# Proportional Fairness in Multi-channel Multi-rate Wireless Networks – Part I: The Case of Deterministic Channels with Application to AP Association Problem in Large-Scale WLAN

Soung Chang Liew, *Senior Member, IEEE* and Ying Jun (Angela) Zhang, *Member, IEEE*

Department of Information Engineering, The Chinese University of Hong Kong
{soung, yjzhang}@ie.cuhk.edu.hk

***Abstract –*** This is Part I of a two-part paper series that studies the use of the proportional fairness (PF) utility function as the basis for resource allocation and scheduling in multi-channel multi-rate wireless networks. The contributions of Part I are threefold. (i) We present the fundamental properties and physical/economic interpretation of PF optimality. We show that PF leads to equal airtime allocation to users for the single-channel case; and equal *equivalent airtime* allocation to users for the multi-channel case. In addition, we also establish the Pareto efficiency of joint-channel PF optimal solution (the formulation of interest to us in this paper), and its superiority over the individual-channel PF optimal solution in that the individual user throughputs of the former are all equal to or greater than the corresponding user throughputs of the latter. (ii) Second, we derive characteristics of joint-channel PF optimal solutions useful for the construction of PF-optimization algorithms. In particular, we show that a PF solution typically consists of many zero airtime assignments when the difference between the number of users $U$ and the number of channels $S$, $|U-S|$, is large. We present several PF-optimization algorithms, including a fast algorithm that is amenable to parallel implementation. (iii) Third, we study the use of PF utility for resource allocation in large-scale WiFi networks consisting of many adjacent wireless LANs. We find that the PF solution simultaneously achieves higher system throughput, better fairness, and lower outage probability with respect to the default solution given by today's 802.11 commercial products. Part II of this paper series extends our investigation to the time-varying-channel case in which the data rates enjoyed by users over the channels vary dynamically over time.

*Key words–* **Proportional fairness, Scheduling, Resource allocation, AP association, WLAN, 802.11, WiFi, Wireless networks.**

---

This work was supported under the Area of Excellence Scheme (Project Number AoE/E-01/99) and the Competitive Earmarked Research Grant (Project Number 414305) established under the University Grant Committee of the Hong Kong Special Administrative Region, China.

## I. INTRODUCTION

Resource allocation is a fundamental problem in communications networks when there are competing demands from users for the network bandwidth. When allocating transmission resource to network users, there is generally a tradeoff between system throughput and fairness. Originally proposed by Kelly in the context of rate control of wired networks [1], maximizing the logarithmic utility function, $y=\sum_{i=1}^{U}\log(T_i)$, where $T_i$ is the throughput of user $i$ and $U$ is the total number of users in the system, yields a good balance between system throughput and fairness. The logarithmic utility is often referred to as the Proportional Fairness (PF) utility.

In this two-part paper series, we examine the use of PF utility in the context of wireless networks, in which multiple users are competing for the access of multiple channels, with users possibly enjoying different data rates on each of the channel. Specifically, each user $i$ enjoys a data rate of $b_{i,k}$ on channel $k$, and the issue is to assign the airtime usage of channel $k$ to each and every of the users to achieve optimal PF utility. In Part I of the paper series, $b_{i,k}$ is deterministic and static, while in Part II, $b_{i,k}$ is generalized to be time-varying. Part I corresponds to the scenario where the time scale of rate adaptation (if any) is relatively large so that $b_{i,k}$ can be considered as quasi-static in the context of resource allocation and scheduling. The theories developed in Part I will serve as building blocks of Part II, which studies the scenario where the time scale of rate adaptation is small relative to that of resource allocation and scheduling.

The problem formulation of Part I is as follows. Let $P_{i,k}$ denote the fraction of airtime of channel $k$ used by user $i$. The optimization problem is

$$\begin{aligned}
&\max_{P_{i,k}} y=\sum_{i=1}^{U}\log\left(\sum_{k=1}^{S}P_{i,k}b_{i,k}\right)\\
&s.t.\quad \sum_{i=1}^{U}P_{i,k}=1 \quad \forall k=1,\cdots,S \\
&\qquad P_{i.k}\ge 0 \qquad \forall i=1,\cdots,U;\ k=1,\cdots,S
\end{aligned} \tag{1}$$

where $S$ is the number of channels, and $U$ is the number of users. A distinguishing feature of our formulation compared with prior work is that we explicitly formulate the fractional airtime usage $P_{i,k}$ into our problem definition. Once $P_{i,k}$ is determined, we can then allocate transmission resource and devise scheduling accordingly. Our formulation allows us to apply the solution to not just scheduling in cellular networks, but also to other problem domains in which multiple users compete for common transmission channels, such as AP association and design of multi-access protocol in wireless local area networks, as articulated in the motivating example below.

***Motivating Example***

The formulation in (1) is quite general and applies to resource allocation and scheduling problems in various settings. Part I focuses on the application domain of large-scale wireless local area networks (WLAN), as illustrated in Fig. 1. There are $U$ wireless stations (STA) and $S$ wireless access points (AP) distributed over a geographical region. Suppose that adjacent APs operate on different frequency channels so that there is no co-channel interference among the WLANs. Then, essentially we have $S$ channels in the system. The data transmission rate enjoyed by STA $i$ if it connects to channel $k$ (AP $k$) is $b_{i,k}$, where $b_{i,k}$ is a function of the signal-to-noise ratio (SNR) with respect to channel $k$ (e.g., in 802.11a, there are eight possible data transmission rates: 6, 9, 12, 18, 24, 36, 48, or 54Mbps).

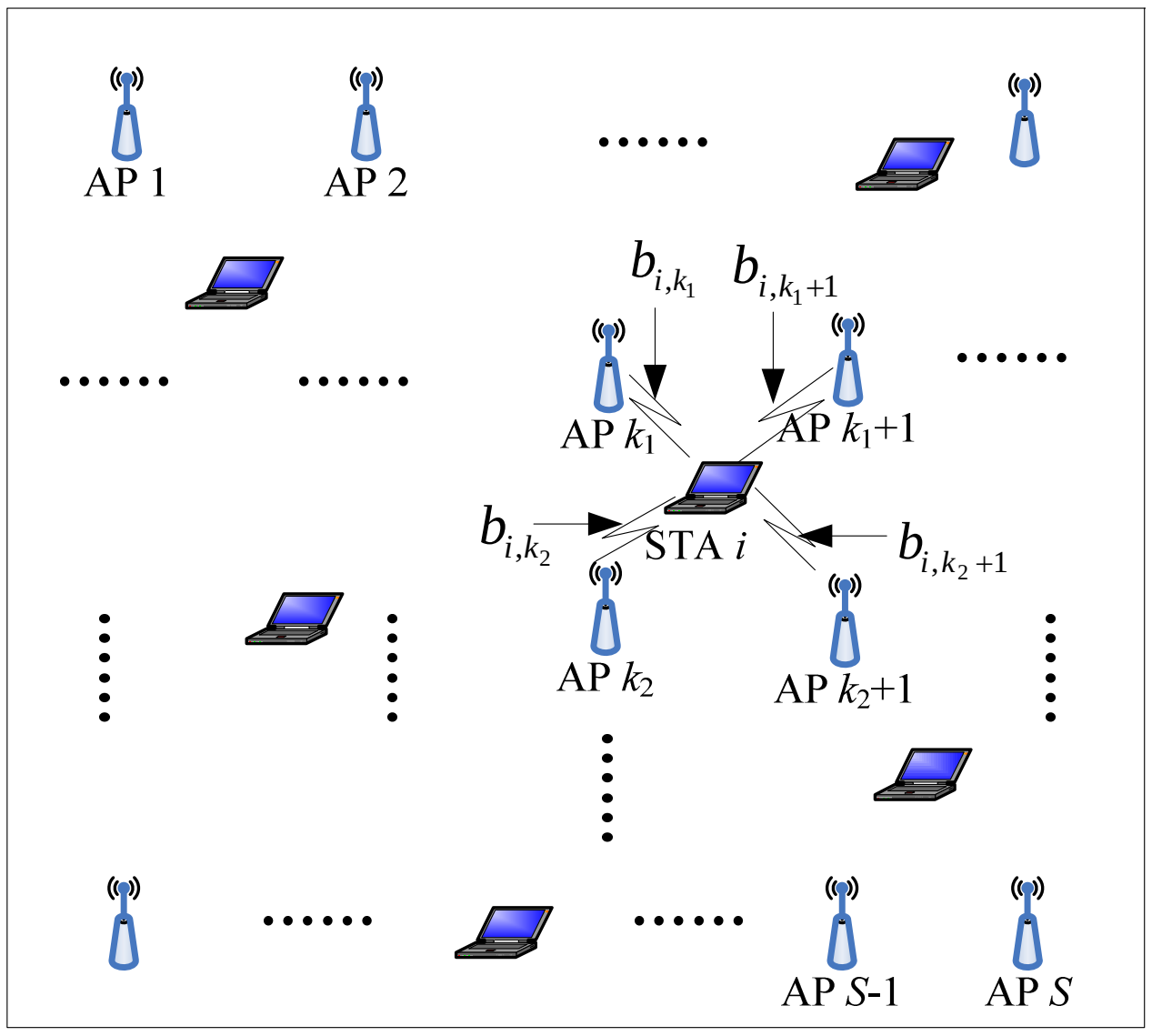

**Fig. 1:** Assigning APs to wireless stations in a wireless LAN

In current 802.11 networks, a STA usually associates itself with the AP with the strongest signal. This may lead to load imbalance and uneven throughputs among STAs when the distribution of STAs is not uniform (e.g., many STAs clustering around one AP, while few STAs near other APs). Formulation (1) does not presume "association with strongest-signal AP". Note that we have assumed the integration of the resource allocation and scheduling problems in the above formulation. In particular, we assume that once airtimes $P_{i,k}$ are determined, there is a medium access control (MAC) scheduling protocol that will make sure that user $i$ uses no more that $P_{i,k}$ fraction of the airtime of channel $k$ when all STAs are busy. For 802.11, this can be achieved by either varying the contention window (CW) or transmission opportunity (TXOP) among the users of the underlying MAC protocol[1] [4].

***Related Work***

The IEEE 802.11 technology today employs a greedy algorithm in medium access in which all stations try to grab as much bandwidth as possible from the network. Ref. [9] provides the definitive analysis on the performance of an 802.11 WLAN in which all stations can hear each other. In a large-scale WLAN containing many APs, such as that discussed in the first motivating example above, not all stations can hear each other. In this situation, throughput distributions among links may be highly uneven and severe unfairness can result under the standard greedy algorithm. A discussion can be found in ref. [10]. Ref. [11] showed that exercise of "offered-load control" so that the source does not behave in a greedy manner can result in better throughput performance in a linear multi-hop network. Ref. [10]-[11] did not treat resource allocation and scheduling within the construct of a mathematical optimization problem. Refs. [12]-[13] are attempts in that direction, assuming the max min utility objective. This paper, in contrast, assumes the PF utility objective. Ref. [4] also touched on the special case of applying PF utility scheduling within a single WLAN (i.e., the single-channel case).

The PF utility has also recently attracted much attention in wireless cellular systems mainly because an opportunistic scheduler implemented in the High Data Rate (HDR) system [15] achieves PF bandwidth sharing among users under certain conditions [2]-[3]. Prior research on PF has mainly focused on systems where there is only a single channel to be assigned. In next-generation wireless networks, users may compete for multiple channels in many scenarios. With the PF utility being $y = \sum_{i=1}^{U} \log(T_i) = \sum_{i=1}^{U} \log\left(\sum_{k=1}^{S} T_{i,k}\right)$ in a multi-channel system, where $T_{i,k}$ is the throughput of user $i$ on channel $k$, the allocation of various channels should be jointly optimized. In "Related Work and Connection with Part I" of the Introduction of Part II, we further articulate the differences of our approach as compared to the prior approaches for wireless cellular systems.

Also in contrast to the previous work is our interest in the characterization of PF optimal solutions. The fundamental properties of PF optimality characterized in this paper series lay down the theoretical groundwork for algorithm design and in-depth performance analysis.

---

[1] On a theoretical basis, $P_{i,k} = c_{i,k} / (1 + \sum_i c_{i,k})$ where $c_{i,k} = CW_{i,k} / TXOP_{i,k}$; the exact formula may need to be modified when other practical considerations are taken into account.

### *Contributions*

Overall, the key contributions of Part I are as follows:

- We provide an economic interpretation for PF utility in wireless networks. We generalize the result in [4] and show that maximizing PF in multi-rate multi-channel wireless networks corresponds to allocating to all users equal *equivalent airtime*, defined as the weighted sum of the airtime of every channel, with the weights being the "value" or "price" of the channels.
- We prove the Pareto efficiency joint-channel PF solution. In addition, let $T_i^*$ be the throughput of user $i$ resulting from (1) and $T_i'$ be the throughput resulting from the individual-channel PF optimization (i.e., wherein the bandwidth of each channel is individually and independently allocated to the users in a PF manner). We prove that $T_i^* \geq T_i'$ for all $i$ if $b_{i,k} > 0$ for all $i$ and $k$. In the Part II, we show that individual-channel PF optimality approaches joint-channel PF optimality under certain conditions.
- We derive the characteristics of PF optimal solutions. We show that a PF solution typically consists of many zero airtime assignments when the difference between the number of users $U$ and the number of channels $S$, $|U-S|$, is large. We take advantage of this property to construct fast $O(S\log S)$ ($O(U\log U)$) algorithms for the 2-user-$S$-channel ($U$-user-2-channel) case.
- Based on the Karush-Kuhn-Tucker (KKT) conditions for multi-channel PF optimality, we design a fast parallel algorithm for the general $U$-user-$S$-channel case.
- We investigate the use of PF and other utility functions for AP assignment in large-scale WiFi networks. We show that the multi-channel PF simultaneously achieves higher system throughput, better fairness, and lower outage probability compared with the default IEEE 802.11 AP association and MAC scheduling scheme in today's commercial products.

The remainder of Part I is organized as follows. Section II provides interpretations of PF optimality in multi-channel wireless networks. Section III gives several characteristics of PF optimal solutions useful for the construction of PF algorithms and interpretation of numerical results later. Section IV presents several PF algorithms. Section V makes use of one of the algorithms to generate numerical results for the application scenario of Fig. 1. Section V concludes Part I of the paper series.

## II. Interpretations of PF Optimality in Multi-Channel Wireless Networks

This section presents the economic interpretations of multi-channel multi-rate PF optimization .

***A. Single-channel Case and Conditions for PF-Optimality in the Multi-channel Case.***

Let us first consider the single-channel case with $S$ = 1. Label the sole channel as channel 1. At any one time, only one user can transmit. We assume that there is enough traffic in the network so that it is always busy. Pick a random point in time. Let $P_{i,1}$ be the probability of finding user $i$ transmitting. The throughput of user $i$ is $T_i = P_{i,1} b_{i,1}$. Differentiating the objective function in (1) with respect to $P_{i,1}$, we get $\partial y / \partial P_{i,1} = 1/P_{i,1}$. The optimal solution is therefore obtained by setting $P_{i,1} = 1/U$ for all $i$.

PF optimality in the single-channel case has a nice and simple interpretation: users should have equal shares of airtime. This makes economic sense in situations where the users are subscribers who pay the same subscription fee to the service provider. In [6], it was shown that a user that transmits at very low rate because of poor SNR can easily drag down the performance of all other users in an 802.11 WLAN, because of the excessive airtime it uses. As a result, everybody suffers because of the "poor" user. With PF scheduling, this problem can be removed, because equal airtime usage establishes a sort of "firewall" among users [4]. One can of course generalize the concept to situations where different users have different priorities (see Extension to Theorem 1 in Section 2.B) and should therefore be allocated different amounts of airtimes. The key concept, however, remains that "airtime" rather than "throughput" should be the resource to be allocated because it is the "currency of fairness": users affect each other directly through the relative portions of airtimes allocated to them.

We shall see that unlike in a single-channel system, PF optimality in a multi-channel system does not mean equal "physical" airtime usage. The airtime of each channel must first be weighted by a "shadow price". Once that is done the concept of an equivalent airtime can then be defined so that PF optimality means equal equivalent airtime among all users. We first present the KKT conditions for the optimization problem of (1) below.

***Karush-Kuhn-Tucker (KKT) Conditions for Multi-channel PF Optimality***

We now turn our attention to the multi-channel problem in (1). Let $[P_{i,k}]$ be the matrix representing a feasible solution, in which rows correspond to users, and columns correspond to channels. Thanks to the concavity of $y$ in the feasible region, the following KKT conditions [7] are necessary and sufficient for a feasible solution $[P_{i,k}^*]$ (with corresponding $T_i^* = \sum_k P_{i,k}^* b_{i,k}$ ) to be optimal:

1. For each channel $k$, for each pair of users $i$ and $j$ with $P_{i,k}^* > 0$ and $P_{j,k}^* = 0$,

$$\left.\frac{\partial y}{\partial P_{i,k}}\right|_{P_{i,k}^*} \geq \left.\frac{\partial y}{\partial P_{j,k}}\right|_{P_{j,k}^*} . \tag{2}$$

That is, $$b_{i,k} \big/ T_i^* \geq b_{j,k} \big/ T_j^* . \tag{3}$$

2. For each channel *k*, for each pair of users *i* and *j* with $P_{i,k}^* > 0$ and $P_{j,k}^* > 0$,

$$\left.\frac{\partial y}{\partial P_{i,k}}\right|_{P_{i,k}^*} = \left.\frac{\partial y}{\partial P_{j,k}}\right|_{P_{j,k}^*} . \tag{4}$$

That is, $$b_{i,k} \big/ T_i^* = b_{j,k} \big/ T_j^* . \tag{5}$$

***2-User-2-Channel Example***

Let $[b_{i,k}]$ be the matrix consisting of the bit rates of different users on different channels. Consider a 2-user-2-channel example in which $[b_{i,k}] = \begin{bmatrix} 1 & 2 \\ 1 & 3 \end{bmatrix}$. It can be verified that the solution $[P_{i,k}^*] = \begin{bmatrix} 1 & 1/4 \\ 0 & 3/4 \end{bmatrix}$ satisfies the KKT conditions and is therefore optimal. We observe the following about the optimal solution: 1) the two users do not have equal airtime on each channel: in fact, user 2 has zero airtime on channel 1; 2) neither are the sums of airtimes on the channels equal: user 1 has total airtime of 1.25, and user 2 has total airtime of 0.75, on the two channels. So, the equal-airtime property of PF optimality in the single channel case does not carry over to the multi-channel case directly.

### *B. Equivalent Airtime in Multi-channel Problem*

In the above 2-user-2-channel example, if we weight the airtime of channel *k* by its "shadow price",

$$\lambda_k^* = \left.\frac{\partial y}{\partial P_{1,k}}\right|_{P_{i,k}^*} = \left.\frac{\partial y}{\partial P_{2,k}}\right|_{P_{2,k}^*} = \frac{b_{1,k}}{T_1^*} = \frac{b_{2,k}}{T_2^*} \tag{6}$$

where $P_{i,k}^* > 0$, then the total weighted airtime of user 1 is $b_{1,1} \big/ T_1^* + 0.25 b_{1,2} \big/ T_1^* = 1$; and the total weighted airtimes of user 2 is $0.75 b_{2,2} \big/ T_2^* = 1$. That is, they are equal. The interpretation is as follows. In the example, since both users can transmit at higher bit rates on channel 2, channel 2 is *more valuable* than channel 1. The shadow price of a channel *k* is a measure of the "value" of the airtime of the channel. Specifically, it is a measure of the potential increase in the utility function *y* for each unit increase in airtime on channel *k* under optimality. We now formally show that the equivalent airtime usage of all users must be equal for multi-channel PF optimality.

***Theorem 1:*** Consider a PF optimal solution $[P_{i,k}^*]$, with shadow price $\lambda_k^*$. Define the equivalent airtime usage of user *i* as $E_i = \sum_k \lambda_k^* P_{i,k}^* = \sum_{k \in K_i^*} \lambda_k^* P_{i,k}^*$ where $K_i^*$ is the subset of channels in which $P_{i,k}^* > 0$. Then, $E_i = 1$ for all *i*.

*Proof:* $E_i = \sum_{k \in K_i^*} \lambda_k^* P_{i,k}^* = (\sum_{k \in K_i^*} b_{i,k} P_{i,k}^*) \Big/ T_i^* = 1$ □

***Extension of Theorem 1:*** Suppose that user *i* is willing to pay a subscription cost of $c_i$, and we modify the utility function in (1) to $y = \sum_i c_i \log\left(\sum_k P_{i,k} b_{i,k}\right)$. Then $E_i = c_i \quad \forall i$.

*Proof:* In this case, the shadow price is $\lambda_k^* = c_i b_{i,k} \Big/ T_k^*$. This gives

$$E_i = \sum_{k \in K_i^*} \lambda_k^* P_{i,k}^* = c_i (\sum_{k \in K_i^*} b_{i,k} P_{i,k}^*) \Big/ T_i^* = c_i \qquad (7)$$ □

We see from the above extension that users may get varying amounts of equivalent airtimes according to the costs they pay.

***C. Pareto Efficiency of PF Optimality***

The PF utility function is just one of many possible utility functions that can serve as the optimization criterion. Within the feasible region defined by the constraints in (1), there are many feasible solutions, each corresponding to one achievable set of throughputs among the users.

***Definition of Pareto Efficiency:*** A feasible solution yielding $T = [T_i]$ is Pareto efficient if and only if one cannot identify another feasible solution yielding $T' = [T_i']$ such that $T_i' \geq T_i$ for all *i*, and there is at least an *i* such that $T_i' > T_i$.

***Assumption:*** In this paper, we make the trivial assumption that any user *i* with $b_{i,k} = 0$ for all *k* would be removed from consideration in the optimization process.

***Theorem 2:*** (a) *Pareto efficiency:* A joint-channel PF-optimal solution is Pareto-efficient. (b) *Superiority of joint-channel optimization over individual-channel optimization:* Suppose that joint-channel PF optimization yields $[T_i^*]$, and individual-channel PF optimization yields $[T_i']$. Then $T_i^* \geq T_i'$ for all *i* if $b_{i,k} > 0$ for all *i* and *k*.

***Proof:*** (a) Suppose the optimal solution yields the utility $y^* = \sum_i \log T_i^*$ and that the solution is not Pareto efficient. Then we can find another solution that yields $y = \sum_i \log T_i$ such that $T_i \geq T_i^*$ for all *i*, and there is at least an $i$ such that $T_i > T_i^*$. In other words, $y > y^*$, and therefore $y^*$ could not have been the optimal solution under PF.

(b) Consider a user $i$. For individual-channel PF optimization, the airtime of each channel is equally shared. So, $T_i' = \sum_k b_{i,k} \big/ U$, where $U$ is the number of users. For joint-channel PF optimization, we have $T_i^* = \sum_k P_{i,k}^* b_{i,k}$. Now, from Theorem 1, we have $E_i = \sum_{k \in K_i^*} \lambda_k^* P_{i,k}^* = \sum_k \lambda_k^* P_{i,k}^* = 1$. This gives $\sum_i \sum_k \lambda_k^* P_{i,k}^* = \sum_k \lambda_k^* = U$. Recall that we have already defined $K_i^*$ to be the subset of channels in which $P_{i,k}^* > 0$. Now, let $\overline{K}_i^*$ be the subset of channels in which $P_{i,k}^* = 0$. Then, according to the KKT conditions, we have $b_{i,k} \big/ T_i^* = \lambda_k^*$ for $k \in K_i^*$ and $b_{i,k} \big/ T_i^* \le \lambda_k^*$ for $k \in \overline{K}_i^*$. Applying this fact in $\sum_k \lambda_k^* = U$, we have $\sum_k b_{i,k} \big/ T_i^* \le U$. This gives $T_i^* \ge \sum_k b_{i,k} \big/ U = T_i'$. □

## III. Characteristics of PF Optimal Solutions

This section discusses characteristics of PF optimal solutions useful for the construction of PF optimization algorithms. We will also use these characteristics to interpret the results in Section V. In the 2-user-2-channel example in Section II, we see that in the optimal solution, one channel is shared and one channel is exclusively assigned to one user. At the same time, one user uses just one channel while the other user uses both channels. It turns out that these results can be generalized. Specifically, in a $U$-user-$S$-channel system, there is an optimal solution in which there are at most $U-1$ shared channels, and there are at most $S-1$ users using more than one channel. These two results are embodied by Theorem 3 and Theorem 4 below, respectively.

### *A. Numbers of Shared and Unshared Channels*

To state Theorem 3 and its corollary more precisely, Let us first define what we mean by "shared channels".

***Definition of Shared Channels:***

1. A channel $k$ is said to be shared if there are at least two non-zero $P_{ik}$, $i \in I_U$, where $I_U$ is the set of all users in the system.
2. A channel $k$ is shared among a subset of $N$ ($N \le U$) users in the system if there are at least two non-zero $P_{ik}$, $i \in I_N$, where $i \in I_N$, and $I_N \subseteq I_U$ is the subset containing the $N$ users.
3. A subset of $N$ ($N \le U$) users are said to share a subset of $M$ ($M \le S$) channels if each and every

of the $M$ channels is shared among the $N$ users. That is, condition 2 above is satisfied for each channel $k$ of the $M$ channels.

***Theorem 3:*** Consider a system with $U$ users and $S$ channels. There is an optimal solution in which the number of shared channels among any $N$ of the $U$ users is no more than $N-1$.

***Corollary 1:*** For a $U$-user-$S$-channel system, there is an optimal solution with no more than $\min(S, U-1)$ shared channels, and with at least $\max(0, S-U+1)$ unshared channels.

***Proof of Corollary 1*:** Obvious from Theorem 3. □

***Proof of Theorem 3*:** See Appendix I. □

***B. Numbers of Users using More than One Channels and Users using just One Channel***

To state Theorem 4 and its corollary more precisely, Let us first define what we mean by "users using more than one channel".

***Definition of Users using More than One Channel:***

1. A user $i$ is said to use more than one channel if there are at least two non-zero $P_{ik}$, $k \in K_S$, where $K_S$ is the set of all channels in the system.
2. A user $i$ is said to use more than one channel in a subset of $M$ ($M \le S$) channels in the system if there are at least two non-zero $P_{ik}$, $k \in K_M$, where $K_M \subseteq K_S$ is the subset containing the $K$ channels.
3. A subset of $N$ ($N \le U$) users are said to use more than one channel in a subset of $M$ ($M \le S$) channels if each and every of the $N$ users uses more than one channel among the $M$ channels. That is, condition 2 above is satisfied for each user $i$ of the $N$ users.

***Theorem 4:*** Consider a system with $U$ users and $S$ channels. There is an optimal solution in which the number of users using more than one channel in any $M$ of the $S$ channels is no more than $M-1$.

***Corollary 2:*** There is an optimal solution with no more than $\min(U, S-1)$ users using more than one channel in the overall system, and with at least $\max(0, U-S+1)$ users using just one channel.

**Comment:** With respect to the AP allocation problem in Fig. 1, to the extent that there are many more STAs than APs, Corollary 2 basically says that most STAs will associate with only one AP.

*Proof of Corollary 2*: Obvious from Theorem 4. □

*Proof of Theorem 4*: See Appendix I. □

## IV. PROPORTIONAL-FAIRNESS ALGORTIHMS

This section presents several algorithms for the PF optimization problem. Subsections A and B consider the special cases where there are only 2 users and 2 channels, respectively. The optimal-solution characteristics derived in the preceding section come in handy for the construction of fast algorithms in these cases. Subsection C presents a parallel algorithm for the general case.

***A. 2-User-S-Channel Case and U-User-2-Channel Case***

We first present a fast $O(S \log S)$ algorithm for the 2-User-*S*-Channel case. The key idea is that the channels can be sorted in such a way that all the channels below a certain boundary channel are used by one user, and all the channels above the boundary channel are used by another user, with at most one channel (the boundary channel) that could possibly be shared. In particular, the idea is to sort the *S* channels according to $b_{1,k} / b_{2,k}$ from large to small. The $O(S \log S)$ computation time is due to the sort operation. Let us relabel the channel numbers according to the sort result so that $b_{1,k} / b_{2,k} \geq b_{1,(k+1)} / b_{2,(k+1)}$ for all *k*.

According to Corollary 1, there is an optimal solution with at most one channel that is shared by the two users. Together with the KKT conditions, this implies the following property:

***Property 1:*** There is an optimal solution with throughputs $T_1^*$ and $T_2^*$, and a channel $S^*$, such that either (i) all channels $k \leq S^*$ are exclusively assigned to user 1; all channels $k > S^*$ are exclusively assigned to user 2; or (ii) channel $S^*$ is shared; all channels $k < S^*$ are exclusively assigned to user 1; all channels $k > S^*$ are exclusively assigned to user 2.

For (i), $$T_1^* / T_2^* \leq b_{1,S^*} / b_{2,S^*} \text{ and } T_1^* / T_2^* \geq b_{1,(S^*+1)} / b_{2,(S^*+1)} . \quad (8)$$

For (ii), $$T_1^* / T_2^* = b_{1,S^*} / b_{2,S^*} . \quad (9)$$

Let us define $T_1^{(k)} = \sum_{l=1}^{k} b_{1,l}$ and $T_2^{(k)} = \sum_{l=k+1}^{S} b_{2,l}$ (i.e., the solution given by exclusively assigning channels 1 to *k* to user 1, and channels *k*+1 to *S* to user 2). Note that

$$T_1^{(k)} / T_2^{(k)} \leq T_1^{(k+1)} / T_2^{(k+1)} \quad \forall k . \quad (10)$$

***Property 2:*** Suppose that $S^l$ and $S^u$ are some known lower and upper bounds for the optimal solution $S^*$ (i.e., $S^l \leq S^* \leq S^u$). Consider a *tentative* solution $S'$ within the bound, in which channels 1 to $S'$ are exclusively assigned to user 1 and the other channels are exclusively assigned to user 2. (i) If $T_1^{(S')} / T_2^{(S')} > b_{1,S'} / b_{2,S'}$, then $S^* \leq S'$. (ii) If $T_1^{(S')} / T_2^{(S')} < b_{1,(S'+1)} / b_{2,(S'+1)}$, then $S^* \geq S'+1$.

To see Property 2(i), consider a channel $k > S'$. Then,

$$\frac{b_{1,k}}{b_{2,k}} \le \frac{b_{1,S'}}{b_{2,S'}} < \frac{T_1^{(S')}}{T_2^{(S')}} \le \frac{T_1^{(k)}}{T_2^{(k)}}. \tag{11}$$

The inequality $\frac{b_{1,k}}{b_{2,k}} < \frac{T_1^{(k)}}{T_2^{(k)}}$ means $k$ cannot be $S^*$ under solution (i) in Property 1. That $\frac{b_{1,k}}{b_{2,k}} < \frac{T_1^{(S')}}{T_2^{(S')}} \le \frac{T_1^{(k)}}{T_2^{(k)}}$ for $k > S'$ means $k$ cannot be $S^*$ under solution (ii) in Property 1 either, since $T_1^* / T_2^* = b_{1,S^*} / b_{2,S^*}$ cannot be achieved by shifting probability from $P_{1,k}$ to $P_{2,k}$.

To see Property 2(ii), consider a channel $k \le S'$. Then,

$$\frac{b_{1,k}}{b_{2,k}} \ge \frac{b_{1,(S'+1)}}{b_{2,(S'+1)}} > \frac{T_1^{(S')}}{T_2^{(S')}} \ge \frac{T_1^{(k)}}{T_2^{(k)}}. \tag{12}$$

The inequality $\frac{b_{1,k}}{b_{2,k}} > \frac{T_1^{(k)}}{T_2^{(k)}}$ means $k$ cannot be $S^*$ under solution (ii) in Property 1 by shifting probability from $P_{1,k}$ to $P_{2,k}$. Also,

$$\frac{b_{1,(k+1)}}{b_{2,(k+1)}} \ge \frac{b_{1,(S'+1)}}{b_{2,(S'+1)}} > \frac{T_1^{(S')}}{T_2^{(S')}} \ge \frac{T_1^{(k)}}{T_2^{(k)}}. \tag{13}$$

So, $k$ cannot be $S^*$ under solution (i) in Property 1 either.

The following is a binary search algorithm to identify $S^*$ based on Properties 1 and 2.

***2-User-S-Channel Algorithm***

Initial solution: $S' \leftarrow \lfloor S/2 \rfloor$; $S^l \leftarrow 1$ ; $S^u \leftarrow S$.

Step 1: $P_{1,k} \leftarrow 1$ for $k = 1,\ldots,S'$; $P_{2,k} \leftarrow 1$ for $k = S'+1,\ldots,S$ ; $P_{i,k} \leftarrow 0$ otherwise.
Compute $T_1^{(S')}$ and $T_2^{(S')}$.

Step 2: if $T_1^{(S')} / T_2^{(S')} > b_{1,S'} / b_{2.S'}$ (*See Property 2(i)*)
then { $S^u \leftarrow S'$; $S' \leftarrow \lfloor (S^l + S^u)/2 \rfloor$;
if $S^u = S^l$ then goto Step 5; else goto Step 1. }

Step 3: if $T_1^{(S')} / T_2^{(S')} < b_{1,(S'+1)} / b_{2,(S'+1)'}$ (*see Property 2(ii)*)
then { $S^l \leftarrow S'+1$; $S' \leftarrow \lfloor (S^l + S^u)/2 \rfloor$;
if $S^u = S^l$ then goto Step 5; else goto Step 1. }

Step 4: $S^* \leftarrow S'$; (*Condition in Property 1(i) satisfied. All channels exclusively assigned.*)
stop.

Step 5: $S^* \leftarrow S'$; *(Channel $S^*$ is shared.)*

$P_{1,k} \leftarrow 1$ for $k = 1,\ldots,S^* - 1$ and $P_{2,k} \leftarrow 1$ for $k = S^* + 1,\ldots,S$;

$$P_{1S^*} \leftarrow \frac{1 + (b_{2,S^*+1} + \ldots + b_{2,S})\big/ b_{2,S^*} - (b_{1,1} + \ldots + b_{1,S^*-1})\big/ b_{1,S^*}}{2};$$

$P_{2S^*} \leftarrow 1 - P_{1S^*}$;

stop.

In a similar vein, we could construct a fast $O(U \log U)$ algorithm for the U-User-2-Channel case. The reader is referred to Appendix II for details of the algorithm.

### *B. U-User-S-Channel Case*

Unfortunately, the simplicity of the 2-User-*S*-Channel and *U*-User-2-Channel algorithms does not carry over to the general *U*-User-*S*-Channel case. The main reason is that with *S* and *U* not equal to 2, a *unique* shared channel and a *unique* user using more than one channel will be lacking (i.e., there could be more than one channel being shared, or more than one user using two or more channels). Thus, the problem cannot be solved by a sorting operation followed by identification of the unique channel or user. We turn to an algorithm of a different spirit for the general *U*-User-*S*-Channel case here. This algorithm is amenable to a parallel implementation. In the algorithm, $P_{i,k}$ is adjusted step by step. In each step, for each channel *k* we try to equate $b_{i,k} / T_{i,k}$ for all users with $P_{i,k} > 0$, so that the KKT condition is satisfied. The computation-intensive steps of the algorithm below (steps marked with *) can be carried out on all channels in parallel for fast execution speed. To start with, we set the initial $P_{i,k}$ to be $1/U$ for all *i* and *k*. The algorithm, however, will work for other initial $P_{i,k}$. To avoid oscillations, we use a factor $\varsigma$ to limit the maximum step size by which $P_{i,k}$ can be adjusted in each iteration.

#### *U-User-S-Channel Algorithm*

Without loss of generality, we focus on an arbitrary channel *k* in the following description.

Initial solution: $P_{i,k} \leftarrow 1/U \ \forall i$; $T_i \leftarrow \sum_{k=1}^{S} b_{i,k} \Big/ U \ \forall i$.

Step 1*: $I_k \leftarrow \{ i \mid P_{i,k} > 0 \}$; $R_k = \sum_{i \in I_k} \frac{b_{i,k}}{T_i} \Big/ |I_k|$.

*(Note that $dy/dP_{i,k} = b_{i,k} / T_i$ and $R_k$ as computed above is the average $dy/dP_{i,k}$ among all users with $P_{i,k} > 0$. The parameter $R_k$ serves as a "reference $dy/dP_{i,k}$" in our algorithm* *such that users with* $dy/dP_{i,k} \geq R_k$ *will have their* $P_{i,k}$ *increased, while users with* $dy/dP_{i,k} < R_k$ *and* $P_{i,k} > 0$ *will have their* $P_{i,k}$ *decreased.)*

Step 2: $I_k^+ \leftarrow \left\{ i \mid i \in I_k \text{ and } \frac{b_{i,k}}{T_i} \geq R_k \right\}$; $I_k^- \leftarrow \left\{ i \mid i \in I_k \text{ and } \frac{b_{i,k}}{T_i} < R_k \right\}$;

$\bar{I}_k^+ \leftarrow \left\{ i \mid P_{i,k} = 0 \text{ and } \frac{b_{i,k}}{T_i} \geq R_k \right\}$ ; $\bar{I}_k^- = \left\{ i \mid P_{i,k} = 0 \text{ and } \frac{b_{i,k}}{T_i} < R_k \right\}$

$R_k^{new} \leftarrow \sum_{i \in I_k \cup \bar{I}_k^+} \frac{b_{i,k}}{T_i} \Big/ \left| I_k \cup \bar{I}_k^+ \right|$;

if $R_k = R_k^{new}$, then goto Step 3

else $R_k \leftarrow R_k^{new}$ and goto Step 2.

(*The purpose of Step 2 is include users with* $P_{i,k} = 0$ *but* $dy/dP_{i,k} \geq R_k$ *(i.e.,* $\bar{I}_k^+$ *) in the set of user whose* $P_{i,k}$ *will be increased. Step 1 included only users with* $P_{i,k} > 0$ *as a first attempt.* Note that $R_k$ is adjusted to be the average $dy/dP_{i,k}$ of the users whose $P_{i,k}$ will be increased. )

Step 3*: $\delta_{i,k} \leftarrow \frac{b_{i,k}}{T_i} - R_k \ \forall i \in I_k \cup \bar{I}_k^+$; $\delta_{i,k} \leftarrow 0 \ \forall i \in \bar{I}_k^-$;

$\alpha \leftarrow \min_{i \in I_k^+ \cup \bar{I}_k^+} \frac{1 - P_{i,k}}{\delta_{i,k}}$; $\beta \leftarrow \min_{i \in I_k^-} \frac{-P_{i,k}}{\delta_{i,k}}$; $c \leftarrow \min(\varsigma, \alpha, \beta)$;

$P_{i,k} \leftarrow P_{i,k} + c\delta_{i,k} \ \forall i$

(*The amount by which* $P_{i,k}$ *will be increased (decreased) is proportional to* $\delta_{i,k}$, *with constant of proporationality c. It is easy to show that* $\sum_i c\delta_{i,k} = 0$ *and* $\sum_i \left(P_{i,k} + c\delta_{i,k}\right) = 1$ *. In other words, the airtime reallocation does not change the total airtime usage.*
*Note that* $\alpha$ *and* $\beta$ *are to ensure the new probability assignment stays between 0 and 1 (i.e.,* $0 \leq P_{i,k} + c\delta_{i,k} \leq 1 \forall i$ *). The parameter* $\varsigma$ *imposes a limit on the adjustment of* $P_{i,k}$ *in each iteration to avoid oscillations.*)

Step 4: $T_i \leftarrow \sum_{k=1}^{S} P_{i,k} b_{i,k} \ \forall i$;

if the KKT condition is satisfied, (i.e., $\frac{b_{i,k}}{T_i} = \frac{b_{j,k}}{T_j}$ if $P_{i,k} > 0$ and $P_{j,k} > 0$ and $\frac{b_{i,k}}{T_i} \geq \frac{b_{j,k}}{T_j}$ if $P_{i,k} > 0$ and $P_{j,k} = 0 \quad \forall k$ ), then stop;

else goto Step 1.

The above *U*-user-*S*-channel algorithm is used to generate numerical results for the study in the next section. A non-parallel version of the program has been written using MATLAB. Alternatively, the built-in functions in MATLAB optimization toolbox based on generic algorithms[2] could be used. However, we find that using the generic algorithms takes exceedingly long computational time even for PF-optimization problems of moderate size, making generating a large number of data points for the numerical study in the next section virtually impossible. In contrast, the computational time is quite manageable with the above $U$-user-$S$-channel algorithm, even for a non-parallel version. It typically takes around 0.6 seconds[3] to converge when there are, for instance, 16 APs and 64 mobile stations in the WiFi network shown in Fig. 1. With this kind of time scale, the algorithm is also suitable for actual field deployment beyond mere numerical studies, since AP allocation and re-allocation are usually not invoked in a frequent manner in typical WLAN-usage scenarios where the users are not highly mobile.

## V. NUMERICAL RESULTS: PF RESOURCE ALLOCATION IN WIFI NETWORKS

We now consider the resource allocation problem in WiFi networks with multiple adjacent WLANs (see Fig. 1). In this study, we assume that there are 16 APs being placed in a square grid. The adjacent APs are separated by 20 meters. A wrap-around method is applied to create a torus topology to eliminate the edge effect: i.e., the rightmost column (top row) is adjacent to the leftmost column (bottom row). A mobile station can transmit at different data rates depending on the SNR. The possible data transmission rates and the corresponding required SNRs are listed in Table I. We further assume a two-ray ground model with path loss exponent of 3 and log-normal shadowing with standard deviation of 6 dB. The average SNR (averaged over shadowing) at the cell boundary is 10 dB. That is, there is a 4 dB shadowing margin for achieving a minimum data rate of 1Mbps.

**Table I:** Minimum SNR required for different data transmission rates

| Data Rate (Mbps) | Minimum required SNR (dB) |
|---|---|
| 0 | $-\infty$ |
| 1 | 6 |
| 6 | 10 |
| 9 | 11 |
| 12 | 12 |
| 18 | 13 |
| 24 | 16 |
| 36 | 19 |
| 48 | 26 |
| 54 | 29 |

[2] MATLAB solves the constrained nonlinear optimization problem using a subspace trust region method for large-scale problems and a sequential quadratic programming method for medium scale problems.
[3] We performed our simulations with MATLAB 7.0 on a Pentium 2GHz machine.

For comparison purposes, besides PF, three conventional AP association schemes, namely, maximum throughput (MT), signal-strength based association with intra-cell throughput fairness (SS-TF), and signal-strength based association with intra-cell airtime fairness (SS-AF), are also simulated. MT aims to maximize the total throughput of the WLAN. Each AP selects among all the STAs those that enjoy the highest data transmission rate to serve. If more than one STA has the same highest rate, equal airtime is assigned to these STAs. This strategy is for benchmarking purposes and is impractical because many STAs would then have zero throughputs. SS-TF is adopted in the current 802.11 networks. The STAs associate themselves with the APs with the strongest signal. Meanwhile, the same throughput is guaranteed for the STAs associated with the same AP. With the standard 802.11 MAC protocol [14], Maxmin fairness is achieved within each cell. As shown in [6], within a cell, the performance of other STAs may be dragged down an STA with poor data rate. SS-AF is similar to SS-TF except that the STAs associated with the same AP are allocated equal airtime. As proved in [4], intra-cell equal airtime allocation leads to PF optimality within a single AP coverage; not PF optimality for the overall network, however.

In the first set of experiments, we assume that the STAs are uniformly distributed in the whole area. Fig. 2 plots total throughput versus number of STAs in the overall network. An interesting observation is that when the number of STAs is small relative to the number of APs, the throughput of PF converges to that of MT. This is because most of the APs are exclusively allocated to just one STA in this case (see Theorem 3 and Corollary 1). To maximize the PF utility, the STA chosen by an AP is the one with the highest throughput, which coincides with MT.

In contrast, when the total number of STAs is much larger than the number of APs, the throughput of PF converges to that of SS-AF. This is also due to the characteristic of PF optimal solutions (see Theorem 4 and Corollary 2). When there are many more STAs than APs, most STAs are associated with only one AP, which is usually the one with the strongest signal strength. Meanwhile, PF optimality leads to equal airtime allocation within each cell, which coincides with SS-AF.

Fig. 2 also indicates that SS-AF and PF outperform SS-TF. Most current WiFi products adopt SS-TF, in which (i) each STA associates with the AP with the highest signal strength; and (ii) the default 802.11 MAC scheduling algorithm is used. An STA at cell boundary has weak SNR and transmits at low data rates. With SS-TF, the throughputs of all STAs will be dragged down by these “weak STAs” [6]. With SS-AF, (ii) is modified to ensure equal airtime for all STAs of an AP [4]. The equal airtime allocation establishes a “firewall” between the strong and weak STAs so that the weak STAs do not eat into the airtime of the strong STAs. We also note that whereas SS-AF is better than SS-TF only when number of STAs is large, PF is better than SS-TF for both small and large numbers of STAs.

In Table II, we compare the fairness performance of the PF scheme with other schemes using the Jain’s fairness index [8]:

$$\left|\sum_{i=1}^{U} T_i\right|^2 \Big/ U\sum_{i=1}^{U}(T_i)^2 , \tag{14}$$

We see that the fairness of MT is significantly worse than the other schemes. Comparatively, PF, SS-TF and SS-AF guarantee much fairer service. In particular, PF achieves consistently better fairness than MT, SS-TF, and SS-AF do.

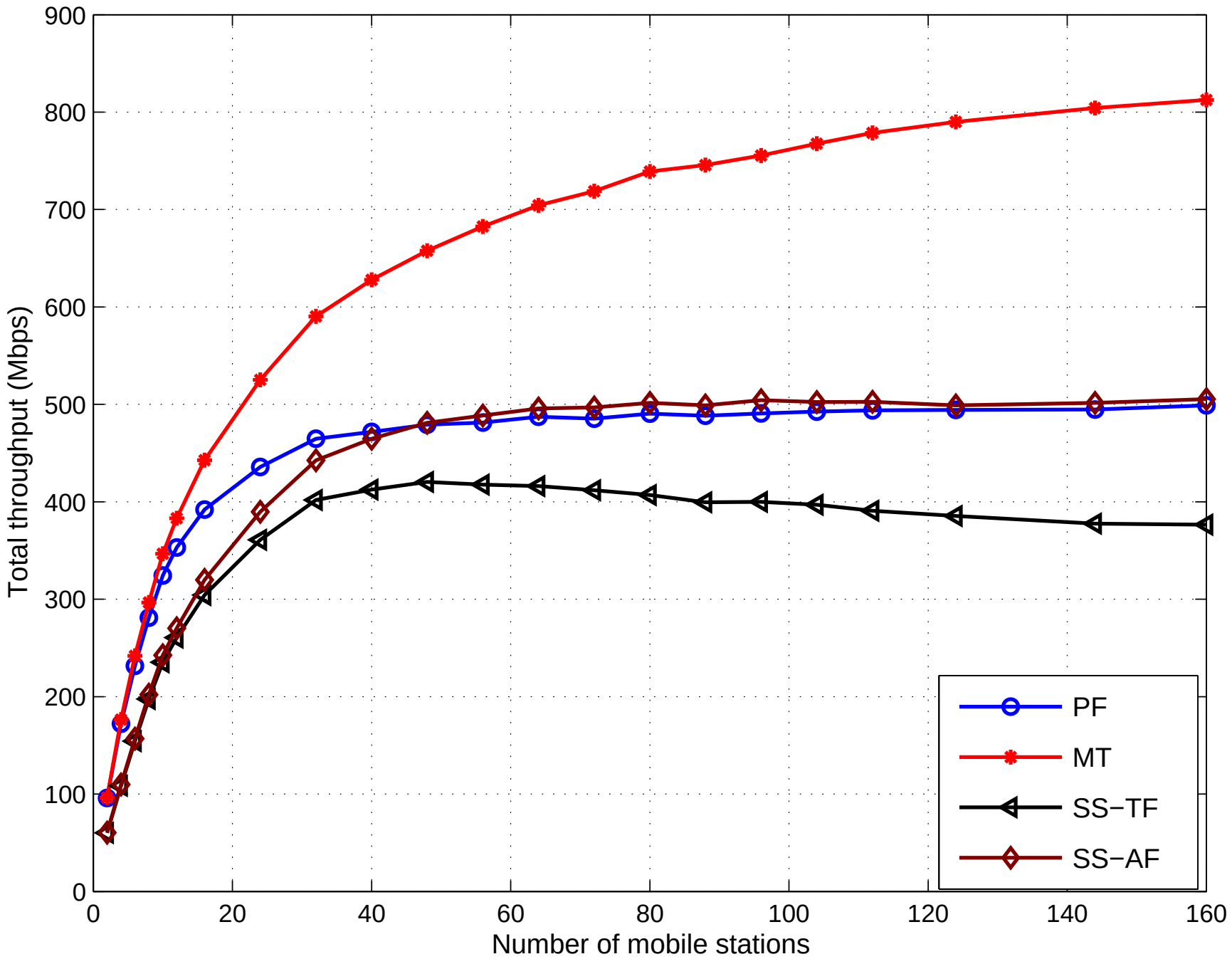

**Fig. 2:** Total Throughput for uniform STA distribution.

**Table II:** Jain’s Fairness index

| | $U$=32 | $U$=48 | $U$=64 |
|---|---|---|---|
| PF | 0.759 | 0.779 | 0.797 |
| MT | 0.432 | 0.291 | 0.277 |
| SS-TF | 0.612 | 0.604 | 0.635 |
| SS-AF | 0.649 | 0.639 | 0.661 |

In Fig. 3, we investigate the outage probability. A user is said to be suffering an outage if its throughput is lower than a minimum data-rate requirement, which is assumed to be 1Mbps in the figure. As the figure shows, PF achieves the lowest outage probability among the four schemes.

In Fig. 2, Fig. 3, and Table II, we have demonstrated that PF strikes a good balance between system throughput and fairness. In the following figures, we show that in a WLAN with hot spots, PF can effectively balance traffic loads among the cells. In this set of experiments, the total number of STAs is 64. Out of the 16 APs, one AP is a hot spot. We define the load percentage of the hot spot to be the percentage of users that are located in the hot spot. The users that are not located in the hot spot are randomly distributed in the other cells. We vary the load percentage of the hot spot from 6.25% (i.e., 1/16, which corresponds to uniform STA distribution) to 100%.

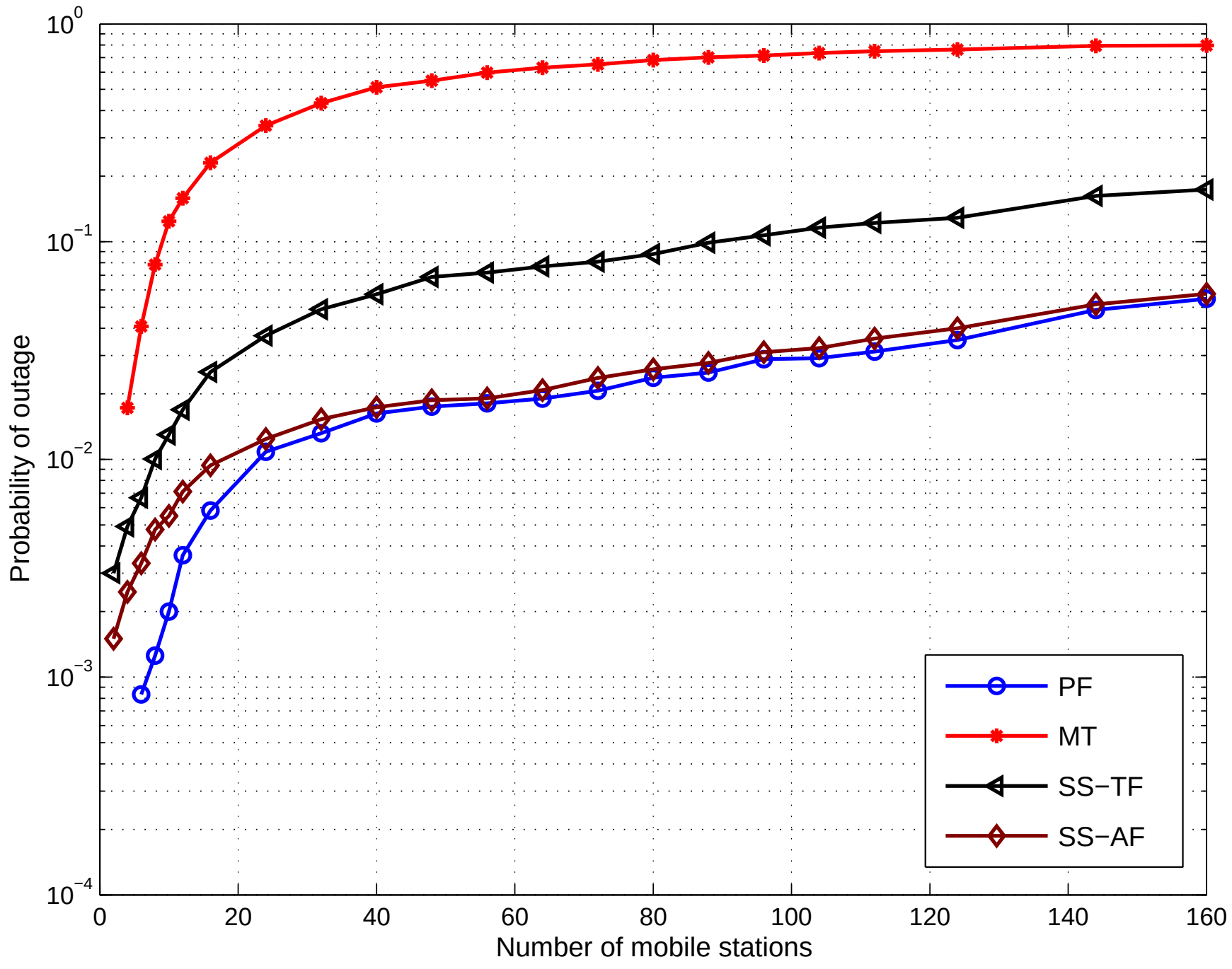

**Fig. 3:** Probability of outage for uniform STA distribution.

A high STA density in the hot spot inevitably results in high outage probability. Fig. 4 shows that PF can mitigate this destructive effect. In particular, unlike the other schemes, its outage probability increases by 3.50% only when the traffic distribution varies from uniform to extremely non-uniform. Fig. 5 illustrates the throughput degradation in the presence of non-uniform traffic distribution. From Fig. 4 and Fig. 5, we can see that PF achieves both higher throughput and lower outage probability compared with SS-TF. Moreover, PF outperforms SS-AF in terms of throughput when the load percentage of the hot spots exceeds 80%. Of course, with respect to MT, PF has lower overall throughput. Such is the price to pay to achieve certain degree of fairness and to prevent starvation on a group of STAs.

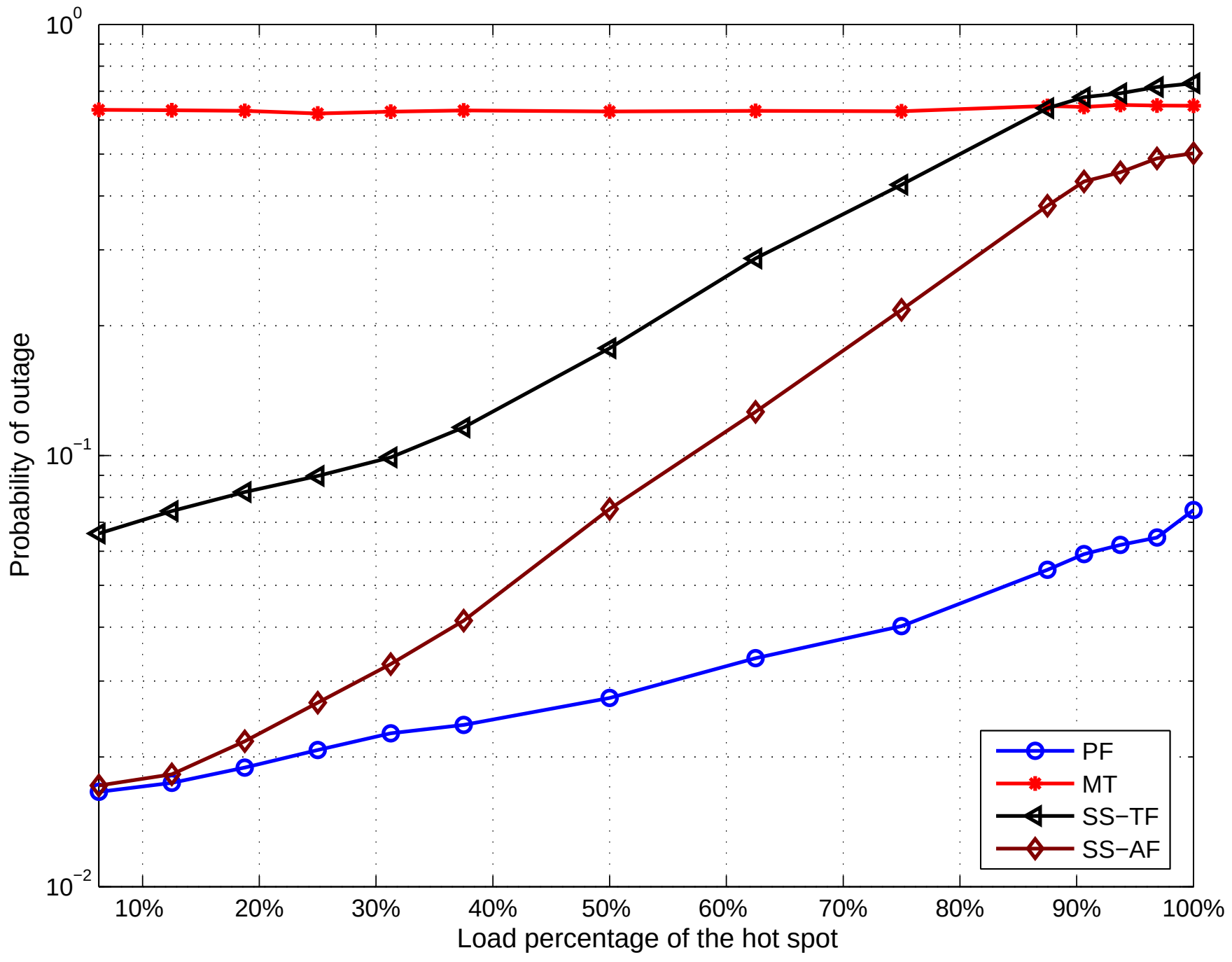

**Fig. 4:** Probability of outage for non-uniform STA distribution. The minimum data-rate requirement is 1Mbps

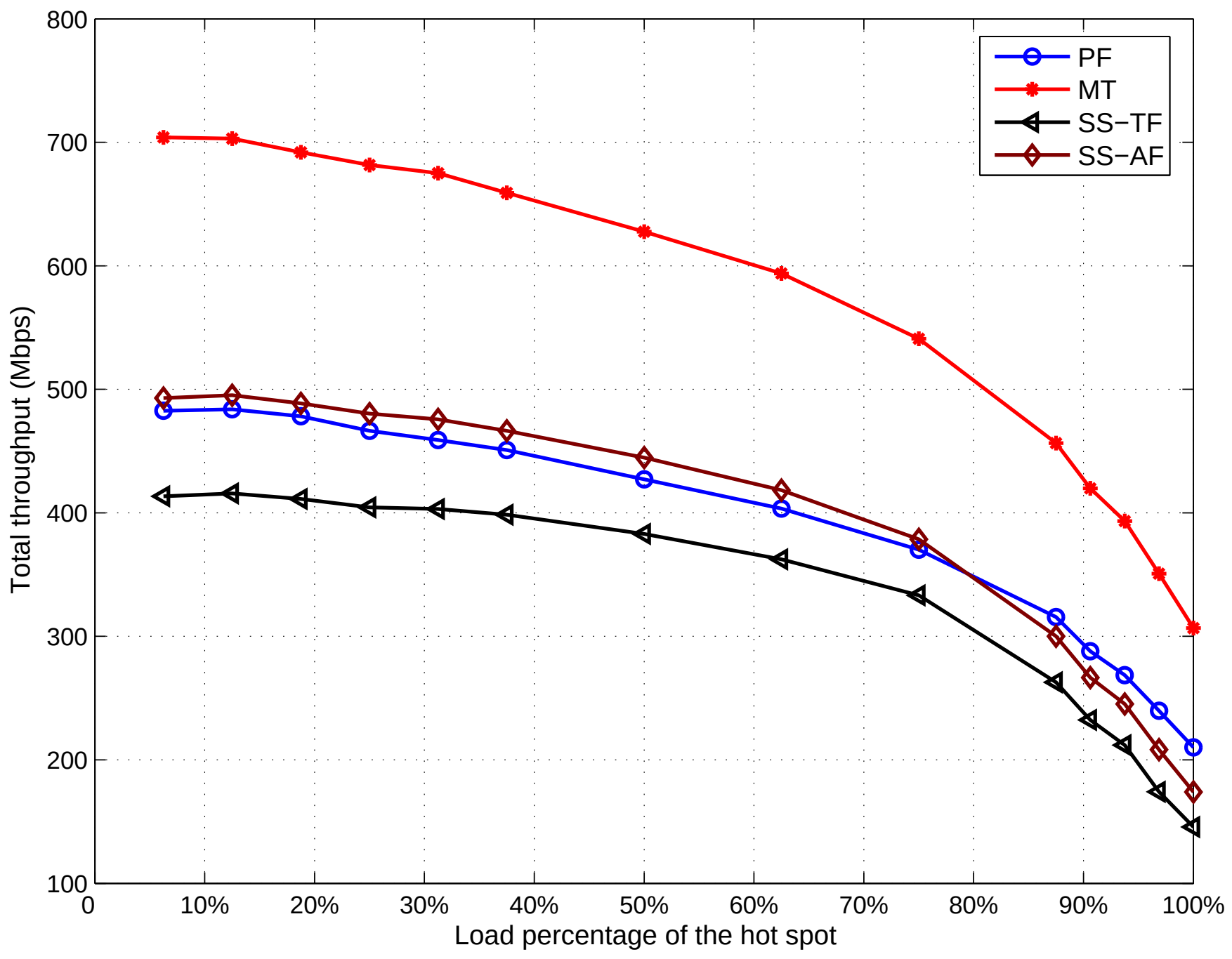

**Fig. 5:** Probability of outage for non-uniform STA distribution

## VI. CONCLUSIONS

This paper has (i) provided economic interpretations for the use the proportional-fairness (PF) utility function for resource allocation in multi-channel multi-rate wireless networks; (ii) derived characteristics of PF optimal solutions and designed several PF algorithms; and (iii) investigated the use of PF and other utilities for resource allocation and AP assignment in large-scale WiFi networks. With regard to (i), we have shown that PF optimization leads to equal *equivalent airtime* allocation to individual users. We have also established the Pareto efficiency of the joint-channel PF-optimal solution and its superiority over the individual-channel PF-optimal solution. With regard to (ii), we show that a PF solution typically consists of many zero airtime assignments when the difference between the number of users *U* and the number of channels *S*, $|U-S|$, is large. We have applied this property to construct fast algorithms for the 2-user-*S*-channel and *U*-user-2-channel cases. In addition, we have presented a fast algorithm amenable to parallel implementation for the general *U*-user-*S*-channel case. With regard to (iii), we have found that using the PF utility function achieves a good balance between system throughput and fairness compared with using the other utility functions. In particular, PF simultaneously achieves higher system throughput, better fairness, and lower outage probability with respect to the default 802.11 AP association and MAC scheduling scheme in today's commercial products. This is the case for uniform as well as non-uniform, and dense as well as sparse, user distributions in the wireless network. The scenario investigated by us is a static network in which the stations are immobile. If the stations in the WiFi network move around (but perhaps slowly), then the data rates of users on different APs will be time-varying. Extending the current formulation for application in such a scenario may be interesting. Although Part II will consider time-varying data rates, the application scenario there is one in which the channel variations are due to fast fading rather than relative distances of stations to different APs.

## APPENDIX I: Proofs of Theorems 3 and 4

This appendix provides the proofs for Theorems 3 and 4.

***Proof of Theorem 3*:** Consider an optimal solution which yields throughput $T_i^*$ for user *i*. Suppose that in this solution, there are *N* users sharing $M \geq N$ channels (see "Definition of Shared Channels" above). We show that we can find another optimal solution such that the *N* users share no more than $N-1$ channels.

Consider the representation of a feasible solution by a bipartite graph in which the left vertices represent the users and the right vertices represent the channels, and in which there is an edge between a vertex *i* on the left and a vertex *k* on the right if $P_{i,k} > 0$. For the bipartite graph to be loop-free, there can be no more than $(N + M - 1)$ edges. The bipartite graph corresponding to the original optimal solution above contains a loop, because according to our definition of shared channels, there are at least $2M \geq N + M$ edges in the original solution.

***Loop-removal Procedure***

We present a procedure that shifts probability assignments to remove loops while maintaining the throughput $T_i^*$ for each user *i*. Consider a loop with *n* left vertices and *n* right vertices. Label the left vertices $i_1$, $i_2$, .., $i_n$; and the right vertices $k_1$, $k_2$, …, $k_n$, with the edges in the loop being $(i_1, k_1)$, $(i_2, k_1)$, $(i_2, k_2)$, $(i_3, k_2)$, …, $(i_n, k_n)$, $(i_1, k_n)$. The KKT conditions for optimality require

$$\begin{aligned} T_{i_1}^* / T_{i_2}^* &= b_{i_1 k_1} / b_{i_2 k_1} \\ T_{i_2}^* / T_{i_3}^* &= b_{i_2 k_2} / b_{i_3 k_2} \\ &\vdots \\ T_{i_{n-1}}^* / T_{i_n}^* &= b_{i_{n-1} k_{n-1}} / b_{i_n k_{n-1}} \end{aligned} \tag{15}$$

and
$$T_{i_n}^* / T_{i_1}^* = b_{i_n k_n} / b_{i_1 k_n} = (b_{i_2 k_1} / b_{i_1 k1})(b_{i_3 k_2} / b_{i_2 k_2})...(b_{i_n k_{n-1}} / b_{i_{n-1} k_{n-1}}) \tag{16}$$

where the right side of (16) is obtained by substitution from (15). Define $d_h = b_{i_h k_{h-1}} / b_{i_h k_h}$ for $h = 2, 3, ..., n$ and $d_1 = b_{i_1 k_n} / b_{i_1 k_1}$. From (16), we have

$$d_1 d_2 d_3 ... d_n = 1 \tag{17}$$

Define
$$c_h = d_1 d_2 ... d_h \tag{18}$$

and $D = \min\left(P_{i_1 k_1} / c_1, P_{i_2 k_1} / c_1, P_{i_2 k_2} / c_2, P_{i_3 k_2} / c_2, ..., P_{i_h k_h} / c_h, P_{i_{h+1} k_h} / c_h, ..., P_{i_n k_n} / c_n, P_{i_1 k_n} / c_n\right)$. (19)

Suppose that $D = P_{i_h k_h} / c_h$ for some *h* (note: the case where $D$ = $P_{i_{h+1} k_h} / c_h$ follows a similar probability-shifting procedure as below except that the + and – signs are reversed.. Then, shift probabilities as follows to obtain a new solution:

$$\begin{gathered} P_{i_h k_h} \leftarrow P_{i_h k_h} - c_h D = 0; \; P_{i_{h+1} k_h} \leftarrow P_{i_{h+1} k_h} + c_h D \geq 0; \\ P_{i_{h+1} k_{h+1}} \leftarrow P_{i_{h+1} k_{h+1}} - c_{h+1} D \geq 0; \; P_{i_{h+2} k_{h+1}} \leftarrow P_{i_{h+2} k_{h+1}} + c_{h+1} D \geq 0; \\ \vdots \\ P_{i_n k_n} \leftarrow P_{i_n k_n} - c_n D \geq 0; \; P_{i_1 k_n} \leftarrow P_{i_1 k_n} + c_n D \geq 0; \\ \vdots \\ P_{i_h k_{h-1}} \leftarrow P_{i_h k_{h-1}} - c_{h-1} D \geq 0 \end{gathered}$$

After applying the above procedure, the change in $T^*_{i_j}$ for $j = 2,...,n$ is

$$-b_{i_j k_j} c_j D + b_{i_j k_{j-1}} c_{j-1} D = 0\,; \tag{20}$$

and the change in $T^*_{i_1}$ is

$$-b_{i_1 k_1} c_1 D + b_{i_1 k_n} c_n D = -b_{i_1 k_1} d_1 D + b_{i_1 k_n} D = 0\,. \tag{21}$$

So, the new solution remains optimal. Furthermore, one edge in the bipartite graph, $(i_h, k_h)$, has been removed.

If there is still a loop remaining, iterate the above procedure until no loop is left. When there is no loop left, there cannot be more than $(N + M - 1)$ edges. For this new solution, let $M_s$ be the number of shared channels and $M_u$ be the number of unshared channels. Each shared channel is associated with at least two edges; and each unshared channel is associated with one edge (note: an unshared channel cannot be associated with zero edge after the above loop-removal procedure because we started out assuming all the $M$ channels are shared, and the loop-free procedure preserves $\sum_{i \in I_N} P_{ik}$ for each channel $k$, so that at least one user is still associated with channel $k$). So, $M_s \le (N + M - 1 - M_u)/2$. This gives $M_s \le N - 1$. □

***Proof of Theorem 4***: Similar to the proof of Theorem 3, we start out by assuming there are $N \ge M$ users that use more than one channel among the $M$ channels in an optimal solution. We then find a loop-free optimal solution using the loop-removal procedure. So, the number of non-zero $P_{ik}$ in the loop-free optimal solution is no more than $(N + M - 1)$. Each user $i$ must have at least one non-zero $P_{ik}$ after the loop-removal procedure, since otherwise $T_i$ cannot be preserved. Let $N_m$ be the number of users using more than one channel and $N_n$ be the number of users using just one channel after the loop-removal procedure. Then,

$$2N_m + N_n \le N + M - 1\,. \tag{22}$$

So, we have $$N_m \le M - 1 \tag{23}$$

after substituting $$N = N_m + N_n\,.$$ □

The key idea is that the users can be sorted in such a way that all the users below a certain boundary user use just one of the channels, and all the users above the boundary user use just the other channel, with at most one users (the boundary user) using both channels.

## APPENDIX II: Algorithm for the *U*-User-2-Channel Case

We present a fast $O(U \log U)$ algorithm for this case. The key idea is that the users can be sorted in such a way that all the users below a certain boundary user use just one of the channels, and all the users above the boundary user use just the other channel, with at most one users (the boundary user) using both channels. In particular, we sort the $U$ users according to $b_{i,1}/b_{i,2}$ from large to small, and relabel the user index so that $b_{i,1}/b_{i,2} \geq b_{(i+1),1}/b_{(i+1),2}$ for all $i$. According to Corollary 2, there is an optimal solution with at most one user using both channels.

***Property 3:*** To achieve proportional fairness, equal airtime should be assigned to the users that are non-multiply assigned to the same channel.

Corollary 2, the KKT conditions, and Property 3 imply the following property:

***Property 4:*** There is an optimal solution with throughputs $T_i^*$, $i = 1,...,U$, and a user $U^*$, such that either (i) all users $i \leq U^*$ use channel 1 only; all users $i > U^*$ use channel 2 only; or (ii) user $U^*$ use both channels; all users $i < U^*$ use channel 1 only; all users $i > U^*$ use channel 2 only.

For (i),
$$\frac{b_{(U^*+1),1}}{b_{(U^*+1),2}} \leq \frac{U^*}{U - U^*} \leq \frac{b_{U^*,1}}{b_{U^*,2}}. \tag{24}$$

For (ii),
$$\frac{U^* - 1}{U - U^* + 1} < \frac{b_{U^*,1}}{b_{U^*,2}} < \frac{U^*}{U - U^*}. \tag{25}$$

(i) in the above is derived from the fact that $T_{U^*}^* / T_{U^*+1}^* \leq b_{U^*,1} / b_{(U^*+1),1}$, $T_{U^*}^* / T_{U^*+1}^* \geq b_{U^*,2} / b_{(U^*+1),2}$, and Property 3 (i.e., $T_{U^*}^* = b_{U^*,1} / U^*$ and $T_{U^*+1}^* = b_{U^*+1,2} / (U - U^*)$. To see (ii), user $U^*$ uses channels 1 and 2 with probabilities $P_{U^*,1}$ and $P_{U^*,2}$, respectively. By computing $T_i^*$ for all $i$, and setting

$$T_{U^*}^* / T_i^* = b_{U^*,1} / b_{i,1} \text{ for } i = 1,...,U^* - 1 \tag{26}$$

and
$$T_{U^*}^* / T_i^* = b_{U^*,2} / b_{i,2} \text{ for } i = U^* + 1,...,U, \tag{27}$$

we can find an expression for $P_{U^*,1}$ and an expression for $P_{U^*,2}$. The requirements of $P_{U^*,1} > 0$ and $P_{U^*,2} > 0$ lead to (ii).

***Property 5:*** Suppose that $U^l$ and $U^u$ are some known lower and upper bounds for the optimal solution $U^*$ (i.e., $U^l \le U^* \le U^u$). Consider a *tentative* solution $U'$ within the bound, in which users 1 to $U'$ use only channel 1 and the other users use only channel 2. (i) If $\frac{b_{U',1}}{b_{U',2}} < \frac{U'}{U-U'}$, then $U^* \le U'$.

(ii) If $\frac{b_{(U'+1),1}}{b_{(U'+1),2}} > \frac{U'}{U-U'}$, then $U^* \ge U'+1$.

To see Property 5(i), consider a user $i \ge U'+1$. Then,

$$\frac{i}{U-i} > \frac{i-1}{U-i+1} \ge \frac{U'}{U-U'} > \frac{b_{U',1}}{b_{U',2}} \ge \frac{b_{i,1}}{b_{i,2}}. \tag{28}$$

So, $i$ cannot be $U^*$ under solutions (i) or (ii) in Property 4. To see Property 5(ii), consider a user $i \le U'$. Then,

$$\frac{i}{U-i} \le \frac{U'}{U-U'} < \frac{b_{(U'+1),1}}{b_{(U'+1),2}} \le \frac{b_{(i+1),1}}{b_{(i+1),2}} \le \frac{b_{i,1}}{b_{i,2}}. \tag{29}$$

So, $i$ cannot be $U^*$ under solutions (i) or (ii) in Property 4 either.

***U-User-2-Channel Algorithm***

Initial solution: $U' \leftarrow \lfloor U/2 \rfloor$; $U^l \leftarrow 1$; $U^u \leftarrow U$.

Step 1: if $\frac{b_{U',1}}{b_{U',2}} < \frac{U'}{U-U'}$ (*see Property 5(i)*)

then { $U^u \leftarrow U'$; $U' \leftarrow \lfloor (U^l + U^u)/2 \rfloor$;

if $U^l = U^u$ then goto Step 4; else goto Step 1}

Step 2: if $\frac{b_{(U'+1),1}}{b_{(U'+1),2}} > \frac{U'}{U-U'}$ (*see Property 5(ii)*)

then { $U^l \leftarrow U'+1$; $U' \leftarrow \lfloor (U^l + U^u)/2 \rfloor$;

if $U^l = U^u$ then goto Step 4; else goto Step 1}

Step 3: $U^* \leftarrow U'$; (*Condition in Property 4(i) satisfied. All users use more than one channel.*)

$$P_{i,1} \leftarrow \begin{cases} \frac{1}{U^*} & \text{for } i = 1, \cdots, U^* \\ 0 & \text{otherwise} \end{cases};\quad P_{i,2} \leftarrow \begin{cases} \frac{1}{U-U^*} & \text{for } i = U^*+1, \cdots, U \\ 0 & \text{otherwise} \end{cases}.$$

stop.

Step 4: $U^* \leftarrow U'$; (*User $U^*$ uses both channels*)

$$P_{i,1} \leftarrow \begin{cases} \dfrac{U-U^*+1}{U} - \dfrac{U^*-1}{U} \cdot \dfrac{b_{U^*,2}}{b_{U^*,1}} & \text{for } i = U^* \\ \dfrac{1-P_{U^*,1}}{U^*-1} & \text{for } i = 1,\cdots,U^*-1 \\ 0 & \text{otherwise} \end{cases}; \quad P_{i,2} \leftarrow \begin{cases} \dfrac{U^*}{U} - \dfrac{U-U^*}{U} \cdot \dfrac{b_{U^*,1}}{b_{U^*,2}} & \text{for } i = U^* \\ \dfrac{1-P_{U^*,2}}{U-U^*} & \text{for } i = U^*+1,\cdots,U \\ 0 & \text{otherwise} \end{cases}$$

stop.